\documentclass[prd,aps,twocolumn,superscriptaddress,nofootinbib,showpacs,preprintnumbers,longbibliography]{revtex4-2}

\usepackage{amsmath}
\usepackage{amssymb}
\usepackage{bm}
\usepackage{multirow}
\usepackage{float}
\usepackage{bbold}
\usepackage{epsfig}
\usepackage{amsmath,amsfonts,amssymb}
\usepackage[T1]{fontenc}
\usepackage[dvipsnames]{xcolor}
\usepackage[colorlinks,linkcolor=Red,anchorcolor=blue,citecolor=BurntOrange,urlcolor=NavyBlue]{hyperref}

\usepackage{graphicx}
\usepackage[american]{babel}
\usepackage{enumerate}
\usepackage{color}
\usepackage{mathtools}
\usepackage{booktabs}
\usepackage{subcaption}
\usepackage{soul}
\usepackage{amsfonts}

\setstcolor{NavyBlue}

\usepackage[capitalise]{cleveref}
\crefname{section}{Sec.}{Secs.}
\crefname{table}{Tab.}{Tabs.}
\crefname{figure}{Fig.}{Figs.}
\crefname{equation}{Eq.}{Eqs.}
\crefname{appendix}{Appendix\ }{Appendix\ }

\newcommand{\QQ}{\left \langle \bar{\tilde{Q}}\tilde{Q} \right \rangle}
\newcommand{\s}{{\tilde{\sigma}}}
\newcommand{\p}{{\tilde{\pi}}}
\newcommand{\Q}{{\tilde{Q}}}
\newcommand{\U}{{\tilde{U}}}
\newcommand{\D}{{\tilde{D}}}
\newcommand{\mean}[1]{\left \langle #1 \right \rangle }

\graphicspath{{figures/}}

\allowdisplaybreaks

\begin{document}

\title{Gravitational-Wave Signatures of Chiral-Symmetric Technicolor}

\author{Hao Yang} 
\email{hyang19@fudan.edu.cn}
\affiliation{Center for Field Theory and Particle Physics \& Department of Physics, Fudan University, 200433 Shanghai, China}

\author{Felipe F.~Freitas}
\email{felipefreitas@ua.pt}
\affiliation{Departamento de F\'{i}sica da Universidade de Aveiro and CIDMA Campus de Santiago, 3810-183 Aveiro, Portugal, EU}

\author{Antonino Marcian\`o}
\email{marciano@fudan.edu.cn}
\affiliation{Center for Field Theory and Particle Physics \& Department of Physics, Fudan University, 200433 Shanghai, China}
\affiliation{Laboratori Nazionali di Frascati INFN, Frascati (Rome), Italy, EU}

\author{Ant\'onio P. Morais} 
\email{aapmorais@ua.pt}
\affiliation{Departamento de F\'{i}sica da Universidade de Aveiro and CIDMA Campus de Santiago, 3810-183 Aveiro, Portugal, EU}

\author{Roman Pasechnik}
\email{Roman.Pasechnik@thep.lu.se}
\affiliation{Department of Astronomy and Theoretical Physics,
Lund University, SE 223-62 Lund, Sweden, EU}

\author{Jo\~ao Viana}
\email{jfvvchico@hotmail.com}
\affiliation{Centro de F\'{i}sica Te\'{o}rica e Computacional, Faculdade de Ci\^{e}ncias,
Universidade de Lisboa, Campo Grande, Edif\'{i}cio C8 1749-016 Lisboa, Portugal, EU}

\begin{abstract}
\vspace{0.5cm}
A chiral-symmetric technicolor model successfully reconciles the tension between electroweak precision tests and traditional technicolor models. Focusing on its simplest realization preserving the conventional Higgs mechanism, we study its primordial gravitational wave signatures originating from first order phase transitions in the early Universe. We found that abundant phase transition patterns arise from a physically viable parameter space. Besides, we have also found gravitational wave signals possibly visible by future experiments, such as LISA, BBO and u-DECIGO. Our results stress the importance of gravitational wave detectors in exploring new physics complementary to ground colliders in the multi-messenger astronomy era.
\end{abstract}

\maketitle


\section{Introduction}
\noindent
The first detection of Gravitational Wave (GW) signatures by the LIGO/Virgo collaboration \cite{LIGOScientific:2016aoc} provided a newly powerful method to study the underlying fundamental physics of our Universe, both at small and large redshift. Particularly, primordial GWs originating from the First Order Phase Transitions (FOPTs) in the early Universe \cite{Kamionkowski:1993fg} are much likely to be observed in the form of stochastic background by the next generation space-based GW interferometers. This important fact leads to gradually increasing attentions on possible probes of particle physics in terms of GW signals \cite{Huang:2016odd,Addazi:2017nmg,Addazi:2019dqt,Wang:2019pet}.

The falsification of the Standard Model (SM) requires the interpretation of the discovery of the Higgs particle at the LHC \cite{ATLAS:2012yve,CMS:2012qbp} as an elementary boson. However, due to the lack of high precision measurements on the properties of the Higgs particle, many New Physics models alternative or extended to the SM Higgs cannot be entirely ruled out. On the other hand, the current SM framework predicts a mild crossover instead of a strong FOPT in the electroweak (EW) phase transitions in the early universe \cite{Kajantie:1996mn,Kajantie:1996qd}, which also causes a difficulty in explaining the baryon asymmetry of our universe \cite{Mazumdar:2018dfl}. Nonetheless, with an additional singlet field this problem can be solved  \cite{Hashino:2018wee,Alves:2018jsw,Kurup:2017dzf,Hashino:2016xoj,Kakizaki:2015wua}. As a result, it turns out to be still essential to study modifications to the well established SM of particle physics.

Historically, the Technicolor (TC) model, featuring an additional strongly coupled sector with confinement, offers a strong substitute to the Higgs mechanism \cite{Weinberg:1975gm,Susskind:1978ms}, where the Higgs field obtains its vacuum expectation value (VEV) via the spontaneous Electroweak Symmetry Breaking (EWSB). Instead, in TC models the EWSB becomes dynamical, being produced by the condensation of technifermions 
$$\QQ \neq 0$$
after the breakdown of the global chiral symmetry. TC models are severely restricted by EW precision experiments \cite{Peskin:1990zt,Peskin:1991sw}. Inspired by the idea of having an alternative to the Higgs mechanism, many variants of TC models have been put forward to fulfill constraints from colliders, e.g.\cite{Appelquist:1986an,Foadi:2007ue,Simmons:1988fu,Kagan:1991gh}. GW signatures of some of these models have been also studied recently \cite{Jarvinen:2009mh,Jarvinen:2010ms,Miura:2018dsy,Azatov:2020nbe}.

In this letter, we focus on a TC variant where a Higgs-like scalar field is added and the conventional Higgs mechanism is effectively preserved in the low-energy limit. This scenario, dubbed as Chiral-Symmetric Technicolor (CSTC) model, features strongly coupled sectors with chiral-symmetric gauge interactions \cite{Pasechnik:2013bxa}. Especially, we consider a minimal techniflavor group $SU(2)_\mathrm{L} \times SU(2)_\mathrm{R}$
breaking to its chiral-symmetric subgroup $SU(2)_\mathrm{L+R}$ at around the EW scale --- see e.g. Ref.~\cite{Pasechnik:2013bxa} for the theoretical introduction to this latter model and its phenomenological implications. Complementary to colliders, potential signatures of the CSTC model in planned GW experiments are the subject of this work.

The paper is organized as follows. In Section \uppercase\expandafter{\romannumeral2}, we revisit the CSTC model and its scalar potential. In Section \uppercase\expandafter{\romannumeral3} GWs originating from FOPTs are introduced. In Section \uppercase\expandafter{\romannumeral4}, we discuss the parameter space of the model and display corresponding GW spectra. Conclusions are summarized in Section \uppercase\expandafter{\romannumeral5}.

\section{The CSTC model}
\noindent
The structure of the CSTC model is inspired by the gauged linear $\sigma$-model \cite{Lee:1967ug,Gasiorowicz:1969kn,Ko:1994en,Urban:2001ru}. Here, considering the simplest case with two Dirac techniflavours in confinement, the global chiral symmetry gets spontaneously broken down to a chiral-symmetric subgroup, subsequently identified as the weak gauge group of the SM, i.e.
\begin{equation}
    SU(2)_\mathrm{L} \times SU(2)_\mathrm{R} \to SU(2)_\mathrm{V \equiv L+R} \equiv SU(2)_\mathrm{W}\,.
\end{equation}
This scheme is analogical to the low-energy effective theory of QCD, where the 3-flavour global chiral group breaks down to a chiral-symmetric (vector) subgroup that can be viewed effectively as indistinguishable from the color group $SU(3)_\mathrm{c}$ in the non-perturbative domain of the theory. For phenomenological purposes it is then convenient to adhere our model as much as possible to the standard QCD, but with the confinement energy scale $\Lambda_{\rm TC}$ chosen to be at a GeV energy scale.

As the simplest realization of the CSTC model, we focus on the first generation of the Dirac technifermion doublet
\begin{equation}
    \Q =  
    \begin{pmatrix}
    \U  \\
    \D
    \end{pmatrix}\,,
\end{equation}
having a constituent mass $m_\Q\sim \Lambda_{\rm TC}$. Besides this, the lightest physical states such as the technisigma $\s$, which comes from a singlet scalar field $S$ in the gauge basis, and the technipions $\p_a$, which originate from a triplet of gauge-basis pseudoscalar fields $P_a$, $a=1,2,3$, are also introduced \cite{Pasechnik:2013bxa}. Then, as the standard Linear $\s$-Model (L$\s$M) suggests, the relevant part of the Lagrangian for the Yukawa interactions can be written as
\begin{equation}
    \mathcal{L}_Y = -\mathcal{Y}_{\rm TC}\bar{\Q}(S + \imath \gamma_5 \tau_a P_a)\Q\,,
    \label{eq:Yukawa}
\end{equation}
where $\tau_a$ denotes the Pauli matrices, and $\mathcal{Y}_{\rm TC}$ stands for the effective coupling constant of $\s$-Model, representing here a free parameter the value of which we will be assume to be in the perturbative regime, i.e. $\mathcal{Y}_{\rm TC} \!< \! \sqrt {4 \pi}$, in the loop-calculation for the effective potential\footnote{The assumption that $\mathcal{Y}_{\rm TC}$ is in a perturbative regime is compatible with a temperatures' domain that fulfills $T \!<\!\!<\! \Lambda_{\rm TC}$. Indeed, at temperatures well below the confinement scale $\Lambda_{\rm TC}$, the conditions for a dilute-gas approximation outside the Fermi-scale can be justified moving from the technimesons' interaction term \eqref{eq:Yukawa}, and the finite-temperature analysis can be assumed to remain in the perturbative regime.}.

The potential (renormalisable) part of the scalar self-interactions reads, in general,
\begin{equation}
    \begin{aligned}
    V_{\mathrm{self}}&(H,S,P) \\
    = &\frac{1}{2}\mu_S^2(S^2+P^2) + \frac{1}{4}\lambda_{\rm TC}(S^2+P^2)^2 \\
    &+ \mu_H^2 H^\dagger H + \lambda_H (H^\dagger H)^2 - \lambda (H^\dagger H)(S^2+P^2)\,,
    \end{aligned}
\end{equation}
and an extra linear source term providing a pseudo-Goldstone mass to the physical technipions 
is also introduced in the potential, i.e.
\begin{equation}
    V_{\mathrm{source}}(S) = \mathcal{Y}_{\rm TC}S\QQ\,.
\end{equation}
Then, the scalar potential of the CSTC model should be written as
\begin{equation}
    V_0(H,S,P) = V_{\mathrm{self}} + V_{\mathrm{source}}\,.
    \label{eq:V_0}
\end{equation}
In the previous expressions, the scalar fields $H$ and $S$ are represented by the following 
expressions,
\begin{equation}
    \begin{aligned}
    H = \dfrac{1}{\sqrt{2}} 
    \begin{pmatrix}
    G + i G^\prime  \\
    \phi_h + h^\prime + i \eta
    \end{pmatrix}\,,	
    \qquad
    S = \phi_\s + s^\prime\,,
    \end{aligned}
\end{equation}
where $h^\prime$, $\eta$, $G$, $G^\prime$, $s^\prime$ are real scalars. The $h^\prime$ and $s^\prime$ fields are quantum fluctuations around the classical background fields $\phi_h$ and $\phi_\s$, which obtain their VEVs in the zero temperature limit, namely, $\phi_h \equiv v = 246\, \mathrm{GeV}$, and $\phi_\s \equiv u$ at $T = 0$. Thus, we have identified the lighter physical CP-even scalar states as the SM-like Higgs boson, with a possible mixing with another field. Despite the Higgs field being an elementary in this simplest scheme, the linear (in $S$) source term in Eq.~(\ref{eq:V_0}) implies a novel quantum-topological origin of the Higgs and $S$ VEVs, $v$ and $u$, connecting them to the technifermion condensate in the near-conformal limit of the theory $\mu_{H,S}^2\to 0$ and, hence, implying a dynamical origin of EWSB in this limit \cite{Pasechnik:2013bxa,Pasechnik:2014ida}. In this work we will focus on the generic case with no-vanishing $\mu_{H,S}^2$, for a more general study.

For the hierarchy of masses of the two CP-even scalar particles, there exist two possibilities: either the lightest scalar state is the technisigma, namely $m_h > m_\s$, or the Higgs boson is the lightest one, i.e. $m_h < m_\s$. Consistently with Ref.~\cite{Pasechnik:2013bxa}, we only consider the latter case in our analysis. The technipions obtain masses through the linear condensate term, while masses of constituent technifermions come from the VEV of technisigma field $S$, which can be expressed as
\begin{equation}
    m_\p^2 = - \frac{\mathcal{Y}_{\rm TC}\QQ}{u},
    \qquad
    m_\Q = \mathcal{Y}_{\rm TC} u\,.
    \label{eq:mass}
\end{equation}
Note that in analogy to the low-energy hadron physics, in what follows we consider the degenerate technifermion masses scenario where $m_\Q \equiv m_\U = m_\D$. Within this scenario, the Higgs-technisigma mixing can be cast as
\begin{equation}
    \tan 2\theta = \frac{4\lambda uv}{2\lambda_{\rm TC} u^2 + m_\p^2 - 2\lambda_H v^2}\,.
\end{equation}
In developing our phenomenological analysis, we use physical parameters as inputs. In other words, additional parameters in the CSTC model are expressed in terms of five independent quantities, i.e.
\begin{equation}
    \qquad m_\s\,, \qquad m_\p\,, \qquad m_\Q\,, \qquad \mathcal{Y}_{\rm TC}\,, \qquad \theta\,.
    \label{eq:inputs}
\end{equation}
to be randomly sampled within certain physically motivated intervals in numerical scans as discussed below.

\section{Gravitational waves from FOPTs}
\noindent
In order to investigate the process of phase transitions in the early Universe, a comprehensive knowledge of the effective potential at finite temperature is necessary. At one-loop level in the weakly-coupled regime of the considered L$\s$M valid below the temperature of the confinement-deconfinement phase transition, the finite-temperature effective potential reads \cite{Coleman:1973jx,Dolan:1973qd,Quiros:1999jp}
\begin{equation}
    V_{\rm eff}(T) = V_0 + V^{(1)}_{\rm CW} + \Delta V(T) + V_{\rm ct}\,,
    \label{eq:V_eff}
\end{equation}
where $V_0$ denotes the tree-level potential, $V^{(1)}_{\rm CW}$ represents the one-loop Coleman-Weinberg potential, $\Delta V(T)$ and $V_{\rm ct}$ are one-loop thermal corrections and the counterterm part respectively. Including Daisy resummation in the effective potential \cite{Parwani:1991gq,Arnold:1992rz} as well, we also take the finite-temperature corrections to the techniquark condensate term $\QQ$ into consideration \cite{Gerber:1988tt,Ioffe:2001bn}. Details can be seen in Appendix.~\ref{sec:A} and \ref{sec:B}.

Gravitational waves produced during FOPTs can be characterized by two key parameters. The first one is the strength of the phase transition $\alpha$, which is associated with the fraction of the released latent heat to the total radiation energy. By the trace anomaly, this parameter is defined as \cite{Hindmarsh:2015qta,Hindmarsh:2017gnf}
\begin{equation}
    \alpha = \frac{1}{\rho_\gamma} \Big[ V_i - V_f - \dfrac{T_p}{4} \Big( \frac{\partial V_i}{\partial T} - 
    \frac{\partial V_f}{\partial T} \Big) \Big] \,,
    \label{eq:alpha}
\end{equation}
where
\begin{equation}
    \rho_\gamma = g_* \frac{\pi^2}{30} T_p^4\,,  \qquad {\rm with} \quad g_* \simeq 112.5 \,,
\end{equation}
denotes the energy density of the radiation medium at the percolation temperature $T_p$ with respect to the number of relativistic degrees of freedom $g_* \simeq 112.5$ \cite{Grojean:2006bp,Leitao:2015fmj,Caprini:2015zlo,Caprini:2019egz}. In the previous definition, $V_i$ and $V_f$ represent the finite-temperature effective potential expressed in Eq.~(\ref{eq:V_eff}) at the initial and the final phase, which are also respectively the symmetric and the broken phase. We emphasize that throughout this analysis we identify the percolation temperature as the one when phase transitions take place. This is a more accurate treatment to estimate the consequent production of GWs \cite{Wang:2020jrd}.

Another crucial parameter is the inverse time-scale of the phase transition $\beta$, which is conventionally written in units of the Hubble parameter $H$ as
\begin{equation}
    \frac{\beta}{H} = T_{p}  \left. \frac{\partial}{\partial T} \left( \frac{\hat{S}_3}{T}\right) \right|_{T_{p}}\,,
    \label{eq:betaH}
\end{equation}
where $\hat{S}_3$ is the Euclidean action. For further details, we refer to Refs.~\cite{Coleman:1977py,Callan:1977pt,Linde:1980tt,Espinosa:2010hh}.\\

Here we only take into account non-runaway bubbles, and discuss the corresponding GW spectra \cite{Caprini:2019egz}. Alternatively, the strength of the phase transition can be evaluated either as the ratio $\Delta v_p / T_p$ or as $\Delta u_p / T_p$, involving respectively the Higgs and technisigma fields. Here,
\begin{equation}
    \Delta v_p = \left| v^f_p - v^i_p \right|\,,
    \qquad
    \Delta u_p = \left| u^f_p - u^i_p \right|\,.
\end{equation}
are differences of VEVs between the initial and final phases for two fields at the percolation temperature $T_p$. While the aforementioned parameter $\alpha$ is helpful with respect to the GW spectrum, $\Delta v_p / T_p$ and $\Delta u_p / T_p$ are quantities commonly adopted within the studies on EW baryogenesis. In what follows we will consider measurements of these three latter quantities at the same time.

 Generally, there are three sources of GWs originating from FOPTs: bubble wall collisions as they form and expand, sound shock wave (SW) produced by the bubble’s violent expansion, and magnetohydrodynamic (MHD) turbulence. However, since the process of bubble wall collisions does not produce GWs in an efficient manner \cite{Hindmarsh:2017gnf,Ellis:2019oqb}, we neglect this contribution in the following. Besides that, we do not consider the impacts of MHD turbulence because of large theoretical uncertainties \cite{Caprini:2019egz}. In fact, since the SW contribution determines the peak frequency and the peak amplitude within the spectrum, to this extent it is sufficient to study this effect only. In the remainder of this work, we will follow the formalism in Ref.~\cite{Caprini:2019egz} for expressions of the GW peak frequency and the peak amplitude, which are functions of aforementioned parameters $\alpha$ and $\beta / H$.

The primordial GWs are redshifted as the universe expands and later form a cosmic gravitational stochastic background. It is quite naturally expected that higher wall velocities of colliding bubbles can lead to more detectable GW signals. In the numerical analysis, with the \texttt{CosmoTransitions} package \cite{Wainwright:2011kj}, we consider supersonic detonations where the GW peak amplitude is maximized by the bubble wall velocity $v_b$ above the Chapman-Jouguet limit. Nevertheless, the numerical deficiency of the calculation of $\hat{S}_3$ in Eq.~(\ref{eq:betaH}) in the \texttt{CosmoTransitions} package has been pointed out in Ref.~\cite{Freitas:2021yng} recently. In our numerical routine, we make use of the same procedure in order to smoothen the action. Note that for some parameter configurations, multi-step phase transitions leading to sequential GWs arise that need careful treatment, see e.g. Ref.~\cite{Morais:2018uou,Morais:2019fnm,Greljo:2019xan,Aoki:2021oez}. And for models with multi-vacua present, multi-step phase transitions can also occur with multi-peak GWs, see e.g. Ref.~\cite{Zhou:2020stj}.

\section{Gravitational wave signatures of the CSTC}
\noindent
As discussed above, we assume that the scale of the new strongly coupled sector is comparable to or exceeds the one of the EW sector in the SM, namely, we adopt an order-of-magnitude estimate, for simplicity, $\Lambda_{\rm TC} \sim v \sim 200 \mathrm{GeV}$, which is three orders of magnitude larger than the conventional QCD energy scale $\Lambda_{\rm QCD} \sim 200 \mathrm{MeV}$. This hierarchy has important implications on the parameter space of the considered theory.

\subsection{Parameter space}
\noindent
In the QCD-like TC paradigm, it is quite natural to expect that characteristic masses of TC particles scale with respect to the corresponding masses of states in standard hadronic physics by the factor $\xi \sim \Lambda_{\rm TC}/\Lambda_{\rm QCD} \gtrsim 1000$. Hence, we set accordingly a lower bound for the masses in (\ref{eq:inputs}), i.e.
\begin{equation}
    m_\p \gtrsim 140 \mathrm{GeV}\,,
    \
    m_\s \gtrsim 500 \mathrm{GeV}\,,
    \
    m_\Q \gtrsim 300 \mathrm{GeV}\,,
\end{equation}
in our numerical study.

One of the phenomenological advantages of the CSTC model, able to fulfil strict EW constraints, comes from its large parameter space, in contrast with conventional TC models. As shown in Ref.~\cite{Pasechnik:2013bxa}, generally the Peskin-Takeuchi (PT) parameters, characterising the complete EW precision tests, are strongly suppressed and weakly rely on all the physical parameters, except for the mixing angle $\theta$. Furthermore, the parameter space is mainly constrained by the $T$ parameter. Indeed, for the degenerate case where $m_\U = m_\D$, which we focus at here, the EW precision measurements impose a rather small value for the $\s-h$ mixing angle. In the following analysis, we set a relatively conservative bound on the Higgs-technisigma mixing, namely, $\left| \cos\theta \right| > 0.85$.

Moreover, the coupling constants in the tree-level potential should be restricted by the tree-level perturbativity. In our numerical study, we limit these latter dimensionless parameters within the range $\left| \lambda_i \right| < 8$ and $\mathcal{Y}_{\rm TC} < 3$, accounting for corrections from both quantum and finite temperature effects.

\subsection{Gravitational waves spectra}
\begin{figure}[H]
\centering
\begin{center}
\includegraphics[width=0.9\linewidth]{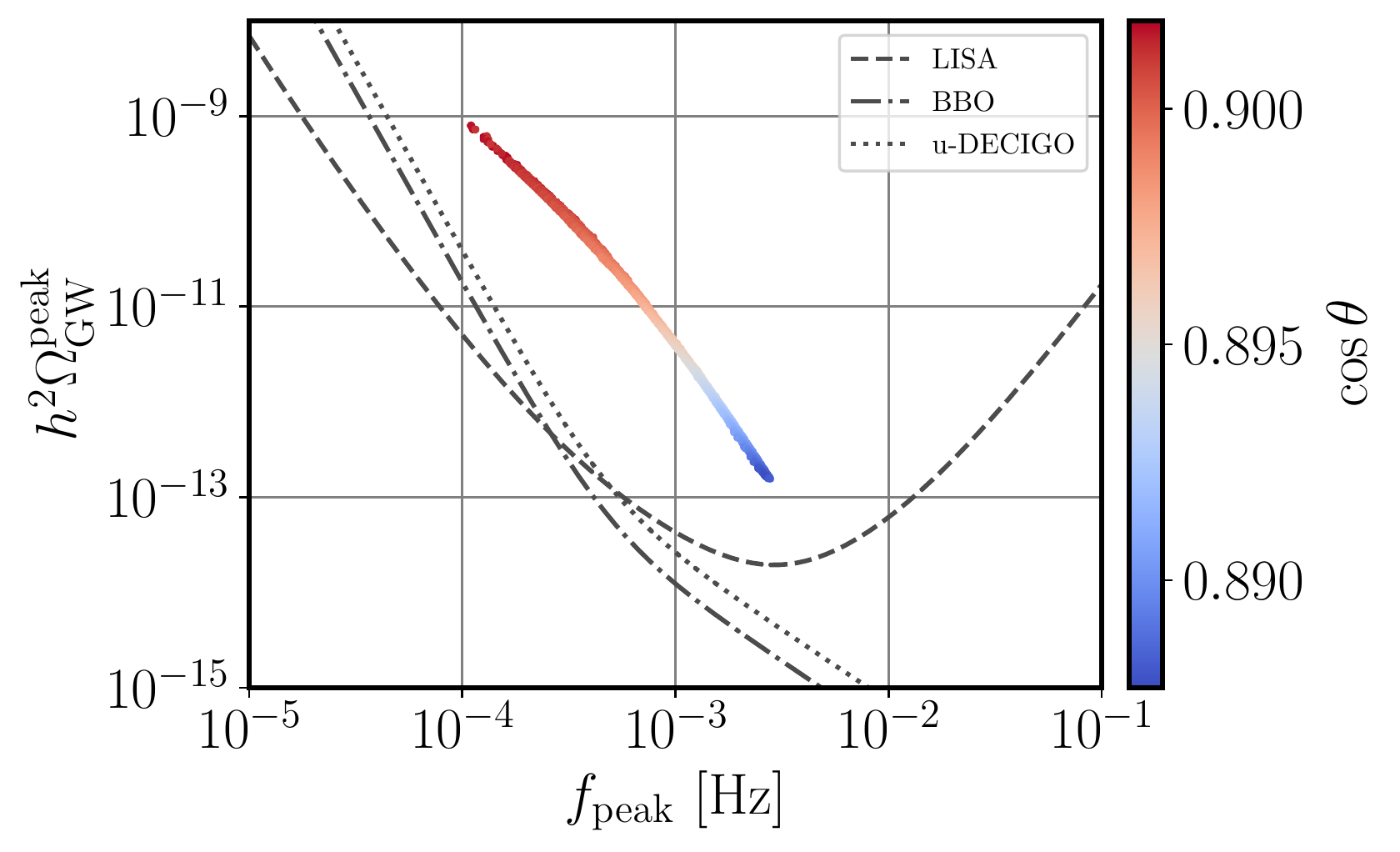}
\end{center}
\vspace{-0.6cm}        
\caption{\footnotesize The GW spectrum as a function of $\cos \theta$, with other model parameters fixed as $m_\s=702.0 \mathrm{GeV}$, $m_\p=347.1 \mathrm{GeV}$, $m_\Q=466.6 \mathrm{GeV}$, $\mathcal{Y}_{\rm TC}=2.86$. Scenarios with small mixing angle $\theta$ can generate observable GW signals. Dashed, dash-dotted and dotted lines represent PISCs for LISA, BBO and u-DECIGO respectively.}
\label{fig:cos}
\end{figure}

\noindent
In order to develop the phenomenology of the CSTC model with GW interferometers, we primarily focus on the parameter space giving rise to signals accessible by future experiments. Fulfilling the restrictions previously discussed, the CSTC model is still capable to generate strong FOPTs in the early universe and leave subsequent visible GWs within the reach of planned missions. In Fig.~\ref{fig:cos} we present the GW spectrum as a function of $\cos \theta$, in order to illustrate the impact of this (most constricted) parameter when all the other parameters are fixed. Fig.~\ref{fig:cos} depicts the peak amplitude of the GW signals $h^2\Omega^{\mathrm{peak}}_{\mathrm{GW}}$ versus its peak frequency $f_{\mathrm{peak}}$ represented in logarithmic scale, with the color bar denoting corresponding values of the parameter $\cos \theta$. Displayed along with result points, within Fig.~\ref{fig:cos} and following figures, there are three grey curves that depict the peak integrated sensitivity curves (PISCs) for sound waves as provided in Ref.~\cite{Schmitz:2020syl}. In Ref.~\cite{Schmitz:2020syl} dashed, dash-dotted and dotted lines represent PISCs of LISA \cite{LISA:2017pwj,Caprini:2015zlo,Caprini:2019egz}, BBO \cite{Corbin:2005ny} and u-DECIGO \cite{Kudoh:2005as} respectively.

Fortunately, we find that observable scenarios favor relatively small, though not vanishing, mixing angle in our results. By specifying one point above PISCs and varying its input $\theta$, it can be seen that gradual approach to the no-mixing limit $\cos^2 \theta \to 1$ enhances the intensity of the peak amplitudes. This characteristic helps detectable signals generated in our parameter space to fulfill the stringent restrictions that arise from collider experiments.

\begin{figure}[H]
\centering
\begin{center}
\includegraphics[width=0.9\linewidth]{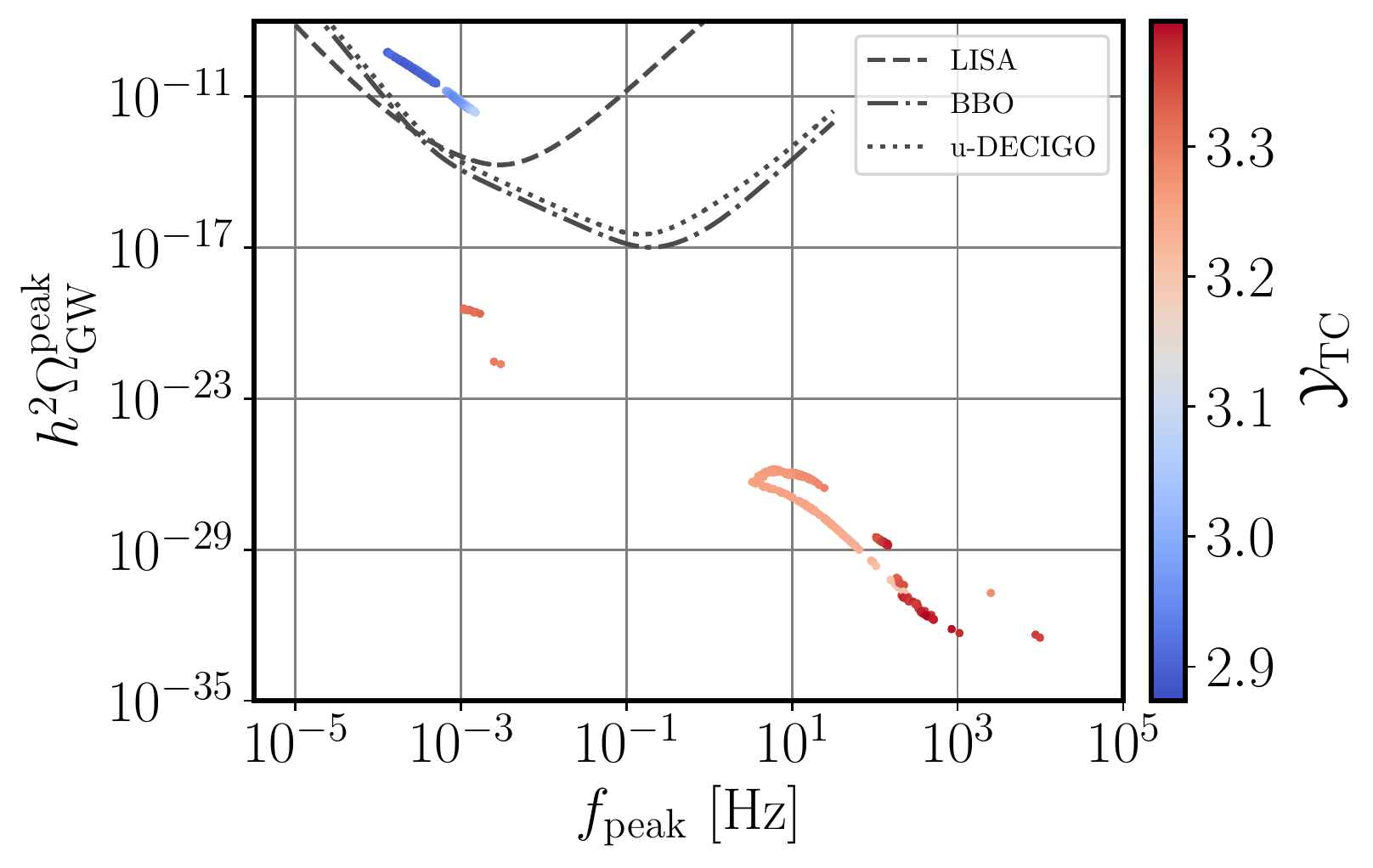}
\end{center}
\vspace{-0.6cm}        
\caption{\footnotesize The GW spectrum as a function of $\mathcal{Y}_{\rm TC}$, with other model parameters fixed as $m_\s=709.3 \mathrm{GeV}$, $m_\p=372.2 \mathrm{GeV}$, $m_\Q=445.0 \mathrm{GeV}$, $\cos \theta=0.904$. A small variation of $\mathcal{Y}_{\rm TC}$ can result in significant variations in corresponding GW spectra.}
\label{fig:gtc}
\end{figure}

On the other hand, the parameter $\mathcal{Y}_{\rm TC}$ is constrained to be small enough in order to avoid non-perturbative effects in the theory. This input serves as a Yukawa-type coupling in the part of the Lagrangian featured in Eq.~(\ref{eq:Yukawa}). In Fig.~\ref{fig:gtc}, we show again the peak amplitude of GW signals $h^2\Omega^{\mathrm{peak}}_{\mathrm{GW}}$ in terms of the corresponding peak frequency, in logarithmic
scale, but this time with the color bar denoting values of the coupling $\mathcal{Y}_{\rm TC}$. It can be seen that, leaving the other inputs unchanged, smaller values than $\mathcal{Y}_{\rm TC} \sim 3.0$, individuating viable regions of the parameter spaces, are able to generate strong FOPTs and result in observable GW signals by future detectors. In general, the strength of GW peak amplitude drops as the value of $\mathcal{Y}_{\rm TC}$ increases. However, a small variation $\sim \mathcal{O}(0.1)$ of $\mathcal{Y}_{\rm TC} \sim 3.0$ can give rise to variations of the corresponding GW spectra as significant as 15 orders of magnitude. In fact, $\mathcal{Y}_{\rm TC}$ between 3.0 and 3.3 cannot generate any FOPTs, while $\mathcal{Y}_{\rm TC} \gtrsim 3.3$ will induce visible ones\footnote{This relatively large gap also guarantees the stability of our results.}. As presented in the figure, starting off within the reach of LISA, the peak amplitudes of GWs shift from the upper-left to the bottom-right, across a large gap in between, and finally end up far below sensitivities of forthcoming experiments. This feature of $\mathcal{Y}_{\rm TC}$ remarkably differs from the other parametric inputs in (\ref{eq:inputs}), the resulting GWs peak signals of which, accordingly, undergo continuous changes.

An additional scalar singlet technisigma mixed with the Higgs field enables the CSTC model to trigger diverse and strong FOPTs in the early Universe, leading to potentially observable GW signals, as in many singlet-scalar extensions of the SM \cite{Hashino:2018wee,Alves:2018jsw,Kurup:2017dzf,Hashino:2016xoj,Kakizaki:2015wua}. This comes from the fact that the additional scalar field $\phi_\s$ will dramatically reshape the finite-temperature effective potential with only the Higgs field present. In fact, apart from one-step FOPTs, we also find two-step and even three-step ones in our physical parameter space, during which two scalar fields are both likely to experience phase transitions with the decrease of the temperature. These configurations will lead to two-peak and even three-peak features in resulting GW spectra. 

\begin{figure}[H]
\centering
\begin{center}
\includegraphics[width=0.9\linewidth]{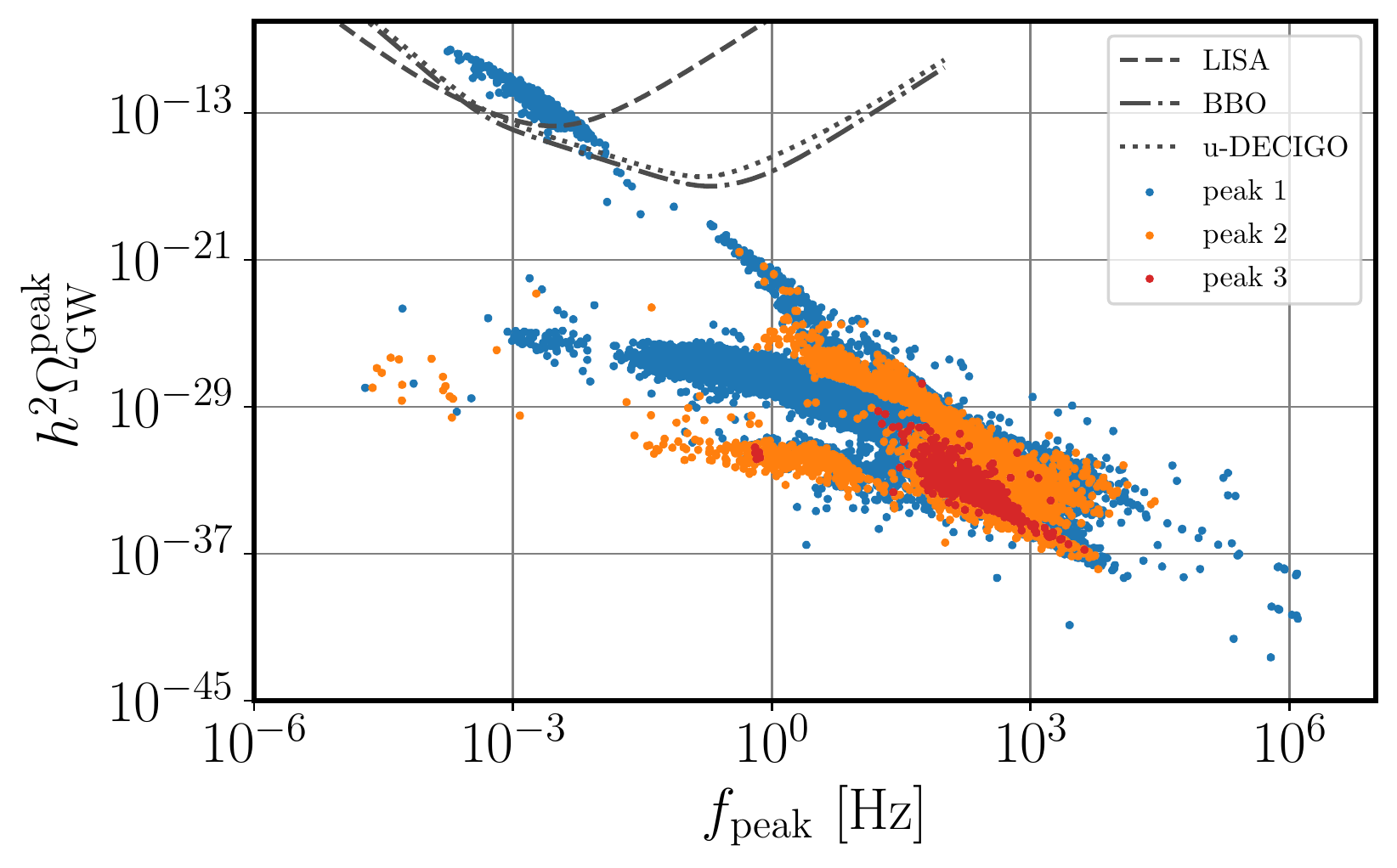}
\end{center}
\vspace{-0.6cm}        
\caption{\footnotesize Scatter plots showing multi-peak features of GW spectra in the CSTC model. The colors represent different peak numbers, 1 (blue), 2 (orange) and 3 (red).}
\label{fig:multipeak}
\end{figure}

Multi-peaks of the GWs spectra appear to be a generic feature of the model we are analysing. We emphasize this phenomenon in Fig.~\ref{fig:multipeak}, where blue, orange and red represent peak 1, peak 2 and peak 3 respectively. A large fraction of the resulting points tends to have double peaks, and in many cases three peaks also arise. As expected, compared to the single-peak scenario, double-peak features are rather few, and similarly triple-peak features. Despite this, in the CSTC model there exists a moderately large parameter space that allows three-step transitions, marked by the red region in the plot. Similarly to what is observed for the single-peak spectra, few multi-peaks spectra fall, for our parameter choices, within the sensitivity that shall be reached by the experiments planned for the forthcoming future. Double-peak and triple-peak spectra are heavily suppressed, leading to difficulty in observing these latter ones in the near future. Even the strongest peak amplitudes of each multi-peak scenarios cannot be detected in the near future. As a result, it is challenging to observe two-peak or three-peak cases in this model and resolve peak diversities in the upcoming measurements.

\begin{figure}[H]
\centering
\begin{center}
\includegraphics[width=0.9\linewidth]{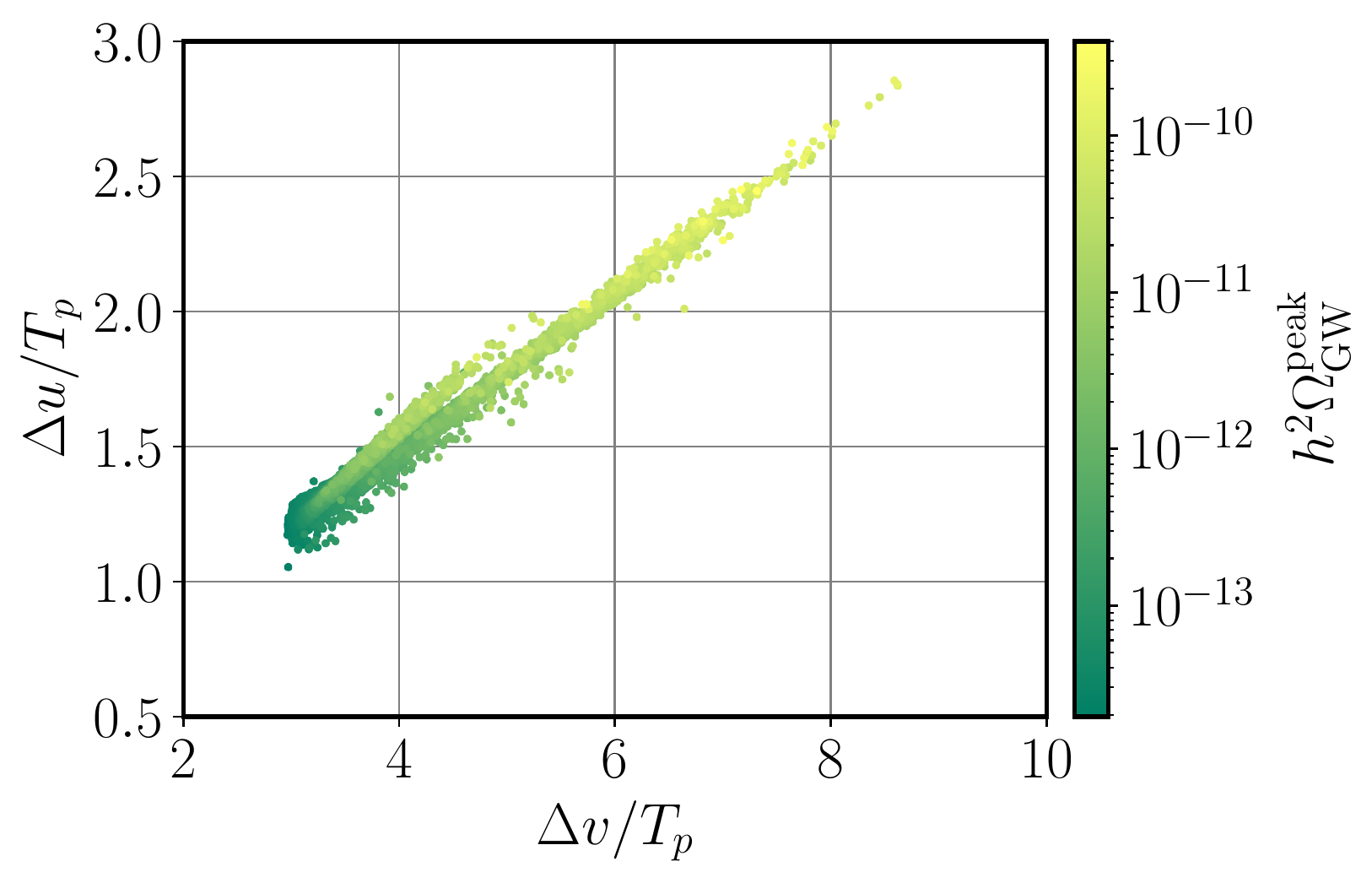}
\end{center}
\vspace{-0.5cm}        
\caption{\footnotesize Scatter plots showing $\Delta v/T_p$ versus $\Delta u/T_p$ of points detectable by LISA, with the color bar denoting the intensity of GW signals. Two branches correspond to two different scans.}
\label{fig:strength}
\end{figure}

In order to study the possibility to verify the CSTC model on satellite interferometers, we now focus on points falling in accessible regions by the LISA mission. In Fig.~\ref{fig:strength} we present $\Delta v/T_p$ as a function of $\Delta u/T_p$ for these points, with the color bar denoting the peak amplitude of the corresponding GWs. Here, two slightly separate branches come from scans in two different intervals of input parameters, which are listed in Tab.~\ref{tab:intervals}. As previously mentioned, $\Delta v/T_p$ and $\Delta u/T_p$ offer alternative measures for the strength of the phase transitions, respectively for EW and chiral phase transitions. We find that for the visible scenarios, $\Delta v/T_p$ varies from 3.0 to 8.6, while $\Delta u/T_p$ ranges between 1.2 to 2.9. At the same order of magnitude, both quantities are larger than one, implying strong phase transitions along the two symmetry-breaking patterns. Noticeably, for a given observable scenario, $\Delta v/T_p$ is at least two times larger than $\Delta u/T_p$, and in those cases in which the GWs signal is strong $\Delta v/T_p$ can even be three times larger than $\Delta u/T_p$. In other words, the EW phase transitions tend to make more contributions to the amplitude of the stochastic gravitational waves background than chiral phase transitions. We conclude that potentially detectable cases favor that both EW and chiral phase transitions are strong first order ones, with a potentially higher contribution from EW ones.

\begin{table}[htb!]
\begin{center}
\begin{tabular}{|c|c|c|c|c|c|}
\hline
                                                                                             No.  & $m_\s$        & $m_\p$        & $m_\Q$        & $\mathcal{Y}_{\rm TC}$  & $\left| \cos\theta \right|$ \\ \hline
\uppercase\expandafter{\romannumeral1} & {[}700,\,750{]} & {[}300,\,350{]} & {[}400,\,450{]} & {[}2.5,\,3.0{]} & {[}0.858,\,0.887{]}           \\ \hline
\uppercase\expandafter{\romannumeral2} & {[}620,\,670{]} & {[}250,\,300{]} & {[}450,\,500{]} & {[}2.2,\,2.8{]} & {[}0.858,\,0.868{]}           \\ \hline
\end{tabular}
\end{center}
\caption{\footnotesize Two different intervals for scanning. Interval \uppercase\expandafter{\romannumeral1} corresponds to the lower branch in Fig.~\ref{fig:strength} and the upper one in Fig.~\ref{fig:SNR}, while Interval \uppercase\expandafter{\romannumeral2} corresponds to the upper (and rather thinner) branch in Fig.~\ref{fig:strength} and the lower one in Fig.~\ref{fig:SNR}. All masses are in units of GeV.}
\label{tab:intervals}
\end{table}

\begin{figure}[H]
\centering
\begin{center}
\includegraphics[width=0.9\linewidth]{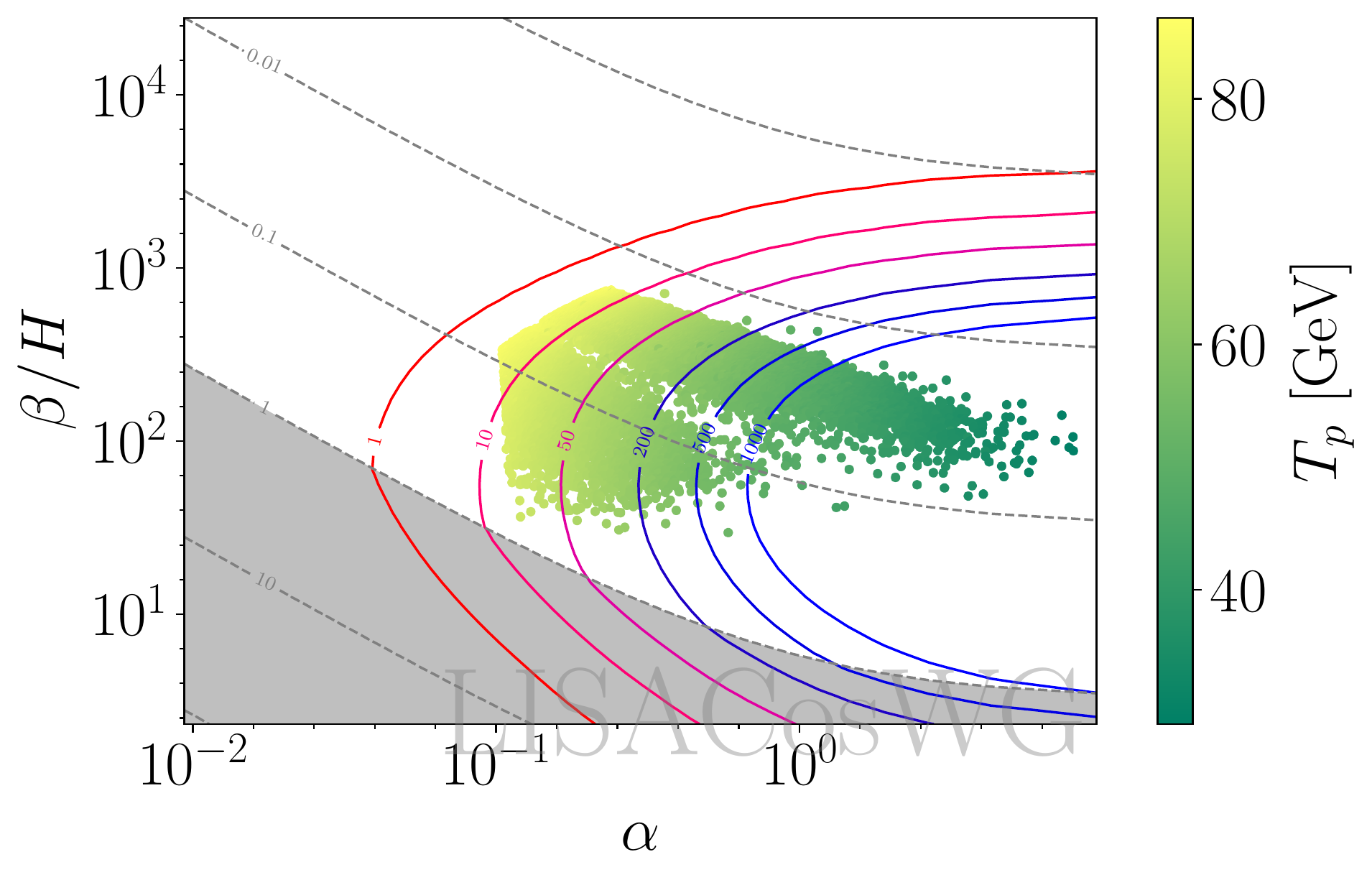}
\end{center}
\vspace{-0.5cm}        
\caption{\footnotesize Scatter plots showing signal-to-noise (SNR) ratio corresponding to points detectable by LISA, with the color bar denoting the percolation temperature. Two branches correspond to two different scans.}
\label{fig:SNR}
\end{figure}

To demonstrate the opportunity to detect those cases we find within LISA reach, we also show the signal-to-noise (SNR) ratio for the mission duration of three years corresponding to these cases in Fig.~\ref{fig:SNR}. Note that results from two intervals in Tab.~\ref{tab:intervals} also form two branches. The colored contours represent the expected SNR values dependent on $T_p$, $g_*$ and $v_b$, while the dashed grey contours display the shock formation time. The grey shaded region highlights where the sound wave treatment is mostly solid, with the acoustic period longer than a Hubble time. Notably, nearly all points detectable by LISA feature a promising SNR more than 10, with a significant fraction of them even more than 1000. This fascinating fact indicates the feasibility to test the CSTC model through the LISA experiment. Last but not least, it is worth mentioning that the peak amplitude, or the SNR, can be slightly amplified for a given combination of a different choice of physical parameters, as is revealed in Tab.~\ref{tab:intervals}.

On the other hand, one may extract from a GW signals event detected the underlying phase transition parameters for the corresponding GW spectra. Besides the bubble wall velocity $v_b$, which is set above the Chapman-Jouguet limit as in supersonic detonations, the percolation temperature $T_p$, the strength of the phase transition $\alpha$ and the inverse time-scale $\beta / H$ are also required to be estimated for GW spectra. From Fig.~\ref{fig:SNR}, we find that in both branches, an increase in $\alpha$ and a decrease in $\beta / H$ will lead to the enhancement of SNR, with a decline of $T_p$ as well. Indeed, this trend of changes in phase transition parameters also results in higher peak amplitudes and smaller peak frequencies. Furthermore, as studied in Ref.~\cite{Hashino:2018wee,Gowling:2021gcy}, there exist remarkable degeneracies in the determination of $T_p$, $\alpha$ and $\beta / H$ for a given GW spectrum. Resorting to the help of the Fisher matrix analysis, one could accurately determine combinations of the aforementioned phase transition parameters and significantly reduce their uncertainties even for values of SNR greater than 20. Our results, featuring large SNR even higher than 1000, are hence capable enough to notably lower their relative uncertainties. In particular, it is safe to assume values of the percolation temperature $T_p< 60$GeV in order to obtain scatter points for the phase transitions consistently with a SNR larger than 50.

\begin{figure}[H]
\centering

\begin{subfigure}{\linewidth}
\centering
\includegraphics[width=0.9\linewidth]{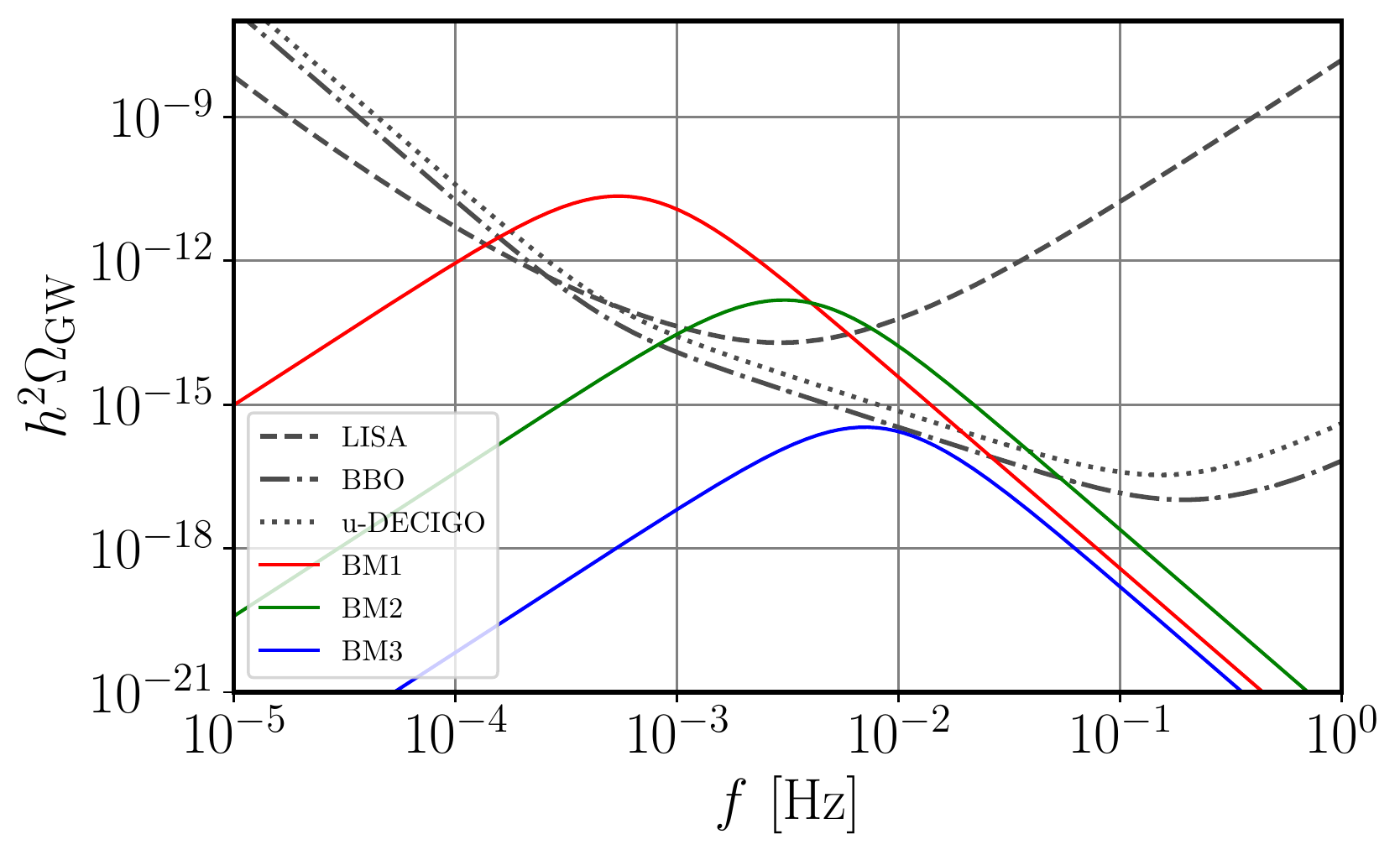}
\caption{\footnotesize Benchmark GW spectra of the CSTC model. The red, green, and blue curves represent benchmark set (BM) 1, 2, and 3 respectively. Phase transition parameters and model parameters are given in Tab.~\ref{tab:phase transition} and Tab.~\ref{tab:model}.}
\label{fig:benchmark}
\end{subfigure}

\begin{subfigure}{\linewidth}
\centering
\includegraphics[width=0.9\linewidth]{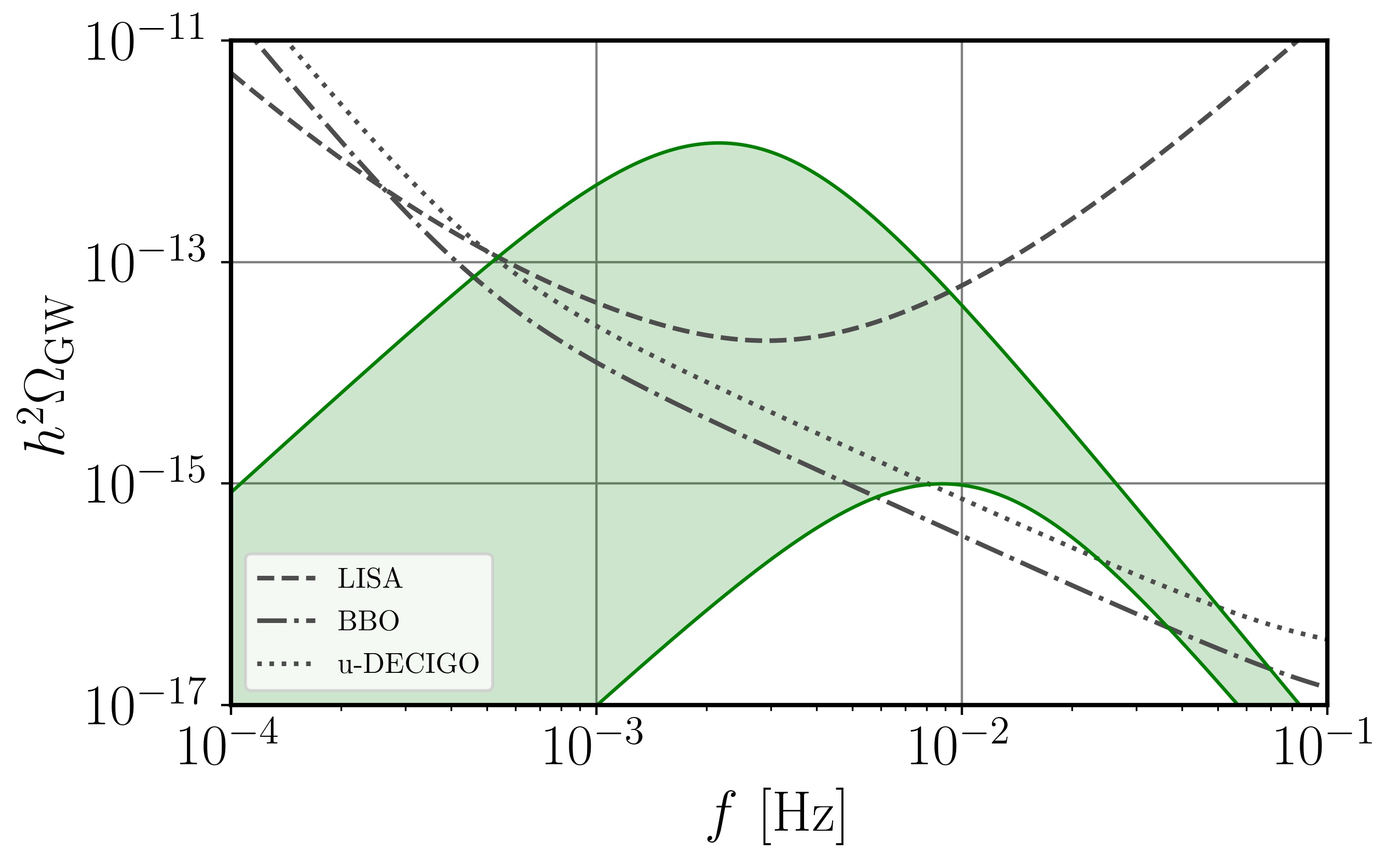}
\caption{\footnotesize The theoretical uncertainty band of the GW spectrum of BM2 in Fig.~\ref{fig:benchmark}. The band shows the variation in the spectrum by varying the renormalization scale by a factor of 3.}
\label{fig:uncertainty}
\end{subfigure}

\caption{\footnotesize Benchmark GW spectra and their representative theoretical uncertainty of the CSTC model. Phase transition parameters and model parameters are given in Tab.~\ref{tab:phase transition} and Tab.~\ref{tab:model}.}
\label{fig:benchmark&uncertainty}
\end{figure}

In order to unveil the specific features of the GW spectra predicted by the CSTC scenario, we choose three benchmark sets and illustrate their GWs intensity spectra in Fig.~\ref{fig:benchmark&uncertainty}, along with the corresponding phase transition parameters shown in Tab.~\ref{tab:phase transition}. For phenomenological purposes, values of the model parameters are presented in Tab.~\ref{tab:model} as well. In Fig.~\ref{fig:benchmark}, benchmark set 1, 2 and 3 are represented by red, green and blue curves respectively, which are all single-peak scenarios. We also present the theoretical uncertainties of our predictions through variations of the renormalization scale in Fig.~\ref{fig:uncertainty}. As discussed in the previous section, even the strongest peak in multi-peak scenarios is inaccessible in the near future, so we decided not to display them. From Fig.~\ref{fig:benchmark}, it is obvious that scenarios with a larger $\alpha$ parameter will retain stronger signals of the GWs spectra. The red curve observable by all the three experiments features $\alpha \sim 1.23$, while the almost hidden blue curve only has $\alpha \sim 0.04$. On the other hand, in spite of different percolation temperatures, both strong EW and chiral phase transitions happen in all visible cases, with ${\Delta v}/{T_p} > 1$ and ${\Delta u}/{T_p} > 1$. In Tab.~\ref{tab:model}, different values of the masses $m_\s$, $m_\p$ and $m_\Q$ are able to induce detectable GWs, under restrictions for small $\mathcal{Y}_{\rm TC}$ and small Higgs-technisigma mixing $\theta$. This fact suggests a large viable parameter space for the phenomenology on space-based interferometers, in complement to the ground-based colliders.

More accurately, there exist indeed large theoretical uncertainties of GW peak amplitudes, mainly coming from the renormalization scale dependence of our finite-temperature effective potential Eq.~(\ref{eq:V_eff}) --- see e.g. \cite{Croon:2020cgk,Gould:2021oba}. Varying the renormalization scale of a factor of 3 in terms of the benchmark set 2 (BM2), the peak amplitude of which is near the limit of LISA, we emphasize the detectability of our model in Fig.~\ref{fig:uncertainty}. Although an overestimated $\mathcal{O}(10^3)$ uncertainty of the peak amplitude is implied, most part of the uncertainty band of our benchmark 2 still falls in observable regions of LISA, BBO and u-DECIGO. Since most of our points lie well within the reach of future missions, this fact validates the potential to test our model in these upcoming experiments.

\begin{table}[htb!]
\begin{center}
\begin{tabular}{|c|c|c|c|c|c|c|}
\hline
                     & Color  & $T_p$  & $\alpha$            & $\beta / H$         & ${\Delta v}/{T_p}$ & ${\Delta u}/{T_p}$  \\ \hline
BM1                  & Red    & 46.36  & 1.23                & 124.50               & 5.47               & 1.86                \\ \hline
BM2                  & Green  & 73.15  & 0.30                & 439.10              & 3.54               & 1.37                \\ \hline
BM3                  & Blue   & 107.10 & 0.04                & 698.24              & 2.36               & 0.98                \\ \hline
\end{tabular}
\caption{Phase transition parameters of three curves in Fig.~\ref{fig:benchmark}. The percolation temperature $T_p$ is given in units of GeV.}
\label{tab:phase transition}
\end{center}
\end{table}

\begin{table}[htb!]
\begin{center}
\begin{tabular}{|c|c|c|c|c|c|c|c|}
\hline
    & Color  & $m_\s$ & $m_\p$ & $m_\Q$ & $\mathcal{Y}_{\rm TC}$ & $\cos \theta$ & $u$   \\ \hline
BM1 & Red    & 785.4  & 239.9  & 591.8  & 2.85     & 0.884         & 207.9 \\ \hline
BM2 & Green  & 744.3  & 303.7  & 470.3  & 2.85     & 0.859         & 165.0 \\ \hline
BM3 & Blue   & 626.2  & 291.1  & 490.5  & 2.38     & 0.859         & 206.4 \\ \hline
\end{tabular}
\caption{Model parameters of three curves in Fig.~\ref{fig:benchmark}. The masses and VEV for the technisigma field $u$ are given in units of GeV.}
\label{tab:model}
\end{center}
\end{table}

\section{Conclusions}
\noindent 
We have studied the possible signatures for gravitational-wave interferometers that are provided by the chiral-symmetric technicolor model, where the global chiral symmetry breaks down to a local chiral-symmetric standard model weak symmetry. Specifically, we have considered a model that accommodates a chiral-symmetric (vector-like) technicolor scenario, involving a new sector of technifermions in confinement that interact with the standard model gauge bosons through vector-like gauge couplings. 
The Higgs boson is accounted for as a separate (fundamental or composite) scalar state, as for the standard model one-doublet. Nonetheless, the electro-weak symmetry breaking can be originated dynamically by the presence of the confined vector-like technifermion sector, induced by the technifermion condensate at the techniconfinement scale, if three orders of magnitude higher than the electro-weak symmetry breaking scale in the nearly conformal limit--- see e.g. \cite{Pasechnik:2013bxa}. Thus, the model encodes an effective standard model Higgs mechanism, complemented with a dynamical electro-weak symmetry breaking. In our work we have considered a more generic case without near-conformal limit.
The model is consistent with electro-weak precision constraints and standard model like Higgs boson observations at the LHC, in the limit of a small Higgs-technisigma mixing \cite{Pasechnik:2013bxa}. The model further predicts at the LHC energy scales the existence of extra new lightest technihadron states, the technipions and the technisigma, responsible for a rich technicolor phenomenology, with detection prospects for the new states, decay modes for technipion and the technisigma and technipion production cross-section previously discussed in \cite{Pasechnik:2013bxa}, in which a physically reasonable regions of the parameter space were individuated.

We have discovered rich patterns of phase transitions induced by this model in the early Universe. More importantly, in the experimentally allowed parameter space, gravitational wave signals observable by forthcoming gravitational interferometers with high SNR can be generated. This allows the potential falsification of (a vast region of the parameter space of) the CSTC model with next-generation experiments, including LISA, BBO and u-DECIGO. In fact, in the multi-messenger astronomy era, information coming from gravitational wave signals will remarkably strengthen the development of particle physics.

\section*{Acknowledgements}
\noindent 
We are indebted with A.~Addazi and Y.F.~Cai for useful correspondence and with Y.F.~Cai for having shared the "LINDA \& JUDY" cluster at USTC. HY wants to thank Xiao Wang for helpful discussions about the percolation temperature. AM wishes to acknowledge support by the Shanghai Municipality, through the grant No.~KBH1512299, by Fudan University, through the grant No.~JJH1512105, the Natural Science Foundation of China, through the grant No.~11875113, and by the Department of Physics at Fudan University, through the grant No.~IDH1512092/001. 
F.F.F.~and A.P.M.~are supported by the Center for Research and Development in Mathematics and Applications (CIDMA) through the Portuguese Foundation for Science and Technology (FCT - Funda\c{c}\~{a}o para a Ci\^{e}ncia e a Tecnologia), references UIDB/04106/2020 and UIDP/04106/2020 and by the projects PTDC/FIS-PAR/31000/2017, PTDC/FIS-AST/3041/2020, CERN/FIS-PAR/0014/2019 and CERN/FIS-PAR/0027/2019. 
A.P.M.~is also supported by national funds (OE), through FCT, I.P., in the scope of the framework contract foreseen in the numbers 4, 5 and 6 of the article 23, of the Decree-Law 57/2016, of August 29, changed by Law 57/2017, of July 19.
R.P.~is supported in part by the Swedish Research Council grant, contract number 2016-05996, as well as by the European Research Council (ERC) under the European Union's Horizon 2020 research and innovation programme (grant agreement No 668679).

\appendix

\section{Thermal Masses}
\label{sec:A}

The $T^2$ term in the high-T expansion of the one-loop thermal corrections $\Delta V(T)$ in the finite-temperature effective potential (\ref{eq:V_eff}) suggests the perturbation theory will break down near the critical temperature. Traditionally this problem is relieved, but not solved, through the performance of an all-order resummation of daisy diagrams \cite{Parwani:1991gq,Arnold:1992rz} . As a result, we need to introduce temperature-dependent corrections to the mass terms in the potential as
\begin{eqnarray}
 \mu_\alpha^2(T) = \mu_\alpha^2 + c_\alpha T^2 \,.
 \label{eq:mu-T}
\end{eqnarray}
In the CSTC model considered here, coefficients of corrections to two mass terms read
\begin{equation}
    \begin{aligned}
    c_H =& \frac{1}{2}\lambda_H - \frac{1}{3}\lambda + \frac{3}{16}g^2 + \frac{1}{16}{g^\prime}^2 \\
    &+ \frac{1}{4}(y_t^2 + y_b^2 + y_c^2 + y_s^2 + y_u^2 + y_d^2) \\ 
    &+ \frac{1}{12}(y_\tau^2 + y_e^2 + y_\mu^2)\,,\\
    c_S =& \frac{1}{2}\lambda_{\rm{TC}} - \frac{1}{3}\lambda + \frac{2}{3}\mathcal{Y}_{\rm TC}^2\,,
    \end{aligned}
\end{equation}
where $g$ and $g'$ are the weak gauge couplings and $y_f$ denotes the Yukawa coupling of the fermion $f$ in the SM.

Besides, in the early universe one needs to consider finite-temperature effects on the whole physical system. Similar to the quark condensate in QCD, we also introduce the finite-temperature corrections to our techniquark condensate. Following the same formulism for hadron physics in Ref.~\cite{Gerber:1988tt,Ioffe:2001bn}, at the leading order we assume the techniquark condensate term similarly takes the finite-temperature form as
\begin{equation}
    \QQ_T = \QQ \left[1 - \frac{1}{4f_\p^2}T^2 - \frac{1}{96f_\p^4}T^4 \right]\,,
\end{equation}
where the technipion decay constant reads
\begin{equation}
    f_\p^2 = - \frac{(m_\U + m_\D)\QQ}{m_\p^2}
\end{equation}
completely analogous with the Gell-Mann-Oakes-Renner relation in QCD.

\section{Renormalization Conditions}
\label{sec:B}
The counterterm part in the finite-temperature effective potential can be written as
\begin{align}
    V_{\rm ct} =& \frac{1}{2}\delta\mu_S^2\phi_\s^2 + \frac{1}{2}\delta\mu_H^2\phi_h^2\\
    &+ \frac{1}{4}\delta\lambda_{\rm TC}\phi_\s^4 + \frac{1}{4}\delta\lambda_H\phi_h^4 - \frac{1}{2}\delta\lambda\phi_h^2\phi_\s^2,
\end{align}

From Eq.~(\ref{eq:V_eff}) one can see that VEVs and physical masses at zero temperature are shifted from their tree-level values by the one-loop CW potential $V^{(1)}_{\rm CW}$. The counterterm potential $V_{\rm ct}$ is hence determined by the requirement that the one-loop effective potential reproduces the same tree-level values at zero temperature \cite{Camargo-Molina:2016moz}. Then renormalization conditions can be expressed as
\begin{equation}
    \begin{aligned}
        \mean{\frac{\partial V_{\rm eff}}{\partial \phi_{\alpha}}}_{\rm vac} = \mean{\frac{\partial V_{0}}{\partial \phi_{\alpha}}}_{\rm vac}\,,&
        \mean{\frac{\partial^2 V_{\rm eff}}{\partial \phi_{\alpha}^2}}_{\rm vac} = \mean{\frac{\partial^2 V_{0}}{\partial \phi_{\alpha}^2}}_{\rm vac}\\
        {\rm and}
        \mean{\frac{\partial^2 V_{\rm eff}}{\partial \phi_{h} \partial \phi_{\s}}}_{\rm vac} &= \mean{\frac{\partial^2 V_{0}}{\partial \phi_{h} \partial \phi_{\s}}}_{\rm vac}\,.
    \end{aligned}
\end{equation}
Consequently, we can solve for counterterms as follows
\begin{align}
    \delta\mu_H^2 =& \frac{1}{2}\mean{\frac{\partial^2 V^{(1)}_{\rm CW}}{\partial \phi_{h}^2}}_{\rm vac} - \frac{3}{2v}\mean{\frac{\partial V^{(1)}_{\rm CW}}{\partial \phi_{h}}}_{\rm vac}\\
    &+ \frac{u}{2v}\mean{\frac{\partial^2 V^{(1)}_{\rm CW}}{\partial \phi_{h} \partial \phi_{\s}}}_{\rm vac},\\
    \delta\mu_S^2 =& \frac{1}{2}\mean{\frac{\partial^2 V^{(1)}_{\rm CW}}{\partial \phi_{\s}^2}}_{\rm vac} - \frac{3}{2u}\mean{\frac{\partial V^{(1)}_{\rm CW}}{\partial \phi_{\s}}}_{\rm vac}\\
    &+ \frac{v}{2u}\mean{\frac{\partial^2 V^{(1)}_{\rm CW}}{\partial \phi_{h} \partial \phi_{\s}}}_{\rm vac},\\
    \delta\lambda_H =& - \frac{1}{2v^2}\mean{\frac{\partial^2 V^{(1)}_{\rm CW}}{\partial \phi_{h}^2}}_{\rm vac} + \frac{1}{2v^3}\mean{\frac{\partial V^{(1)}_{\rm CW}}{\partial \phi_{h}}}_{\rm vac},\\
    \delta\lambda_{\rm TC} =& - \frac{1}{2u^2}\mean{\frac{\partial^2 V^{(1)}_{\rm CW}}{\partial \phi_{\s}^2}}_{\rm vac} + \frac{1}{2u^3}\mean{\frac{\partial V^{(1)}_{\rm CW}}{\partial \phi_{\s}}}_{\rm vac},\\
    \delta\lambda =& \frac{1}{2uv} \mean{\frac{\partial^2 V^{(1)}_{\rm CW}}{\partial \phi_{h} \partial \phi_{\s}}}_{\rm vac}.
\end{align}
\\

\bibliographystyle{spphys}
\bibliography{biblio}

\begin{thebibliography}{10}
\providecommand{\url}[1]{{#1}}
\providecommand{\urlprefix}{URL }
\expandafter\ifx\csname urlstyle\endcsname\relax
  \providecommand{\doi}[1]{DOI \discretionary{}{}{}#1}\else
  \providecommand{\doi}{DOI \discretionary{}{}{}\begingroup
  \urlstyle{rm}\Url}\fi

\bibitem{LIGOScientific:2016aoc}
B.P. Abbott, et~al., Phys. Rev. Lett. \textbf{116}(6), 061102 (2016).
\newblock \doi{10.1103/PhysRevLett.116.061102}

\bibitem{Kamionkowski:1993fg}
M.~Kamionkowski, A.~Kosowsky, M.S. Turner, Phys. Rev. D \textbf{49}, 2837
  (1994).
\newblock \doi{10.1103/PhysRevD.49.2837}

\bibitem{Huang:2016odd}
F.P. Huang, Y.~Wan, D.G. Wang, Y.F. Cai, X.~Zhang, Phys. Rev. D \textbf{94}(4),
  041702 (2016).
\newblock \doi{10.1103/PhysRevD.94.041702}

\bibitem{Addazi:2017nmg}
A.~Addazi, Y.F. Cai, A.~Marciano, Phys. Lett. B \textbf{782}, 732 (2018).
\newblock \doi{10.1016/j.physletb.2018.06.015}

\bibitem{Addazi:2019dqt}
A.~Addazi, A.~Marcian\`o, A.P. Morais, R.~Pasechnik, R.~Srivastava, J.W.F.
  Valle, Phys. Lett. B \textbf{807}, 135577 (2020).
\newblock \doi{10.1016/j.physletb.2020.135577}

\bibitem{Wang:2019pet}
X.~Wang, F.P. Huang, X.~Zhang, Phys. Rev. D \textbf{101}(1), 015015 (2020).
\newblock \doi{10.1103/PhysRevD.101.015015}

\bibitem{ATLAS:2012yve}
G.~Aad, et~al., Phys. Lett. B \textbf{716}, 1 (2012).
\newblock \doi{10.1016/j.physletb.2012.08.020}

\bibitem{CMS:2012qbp}
S.~Chatrchyan, et~al., Phys. Lett. B \textbf{716}, 30 (2012).
\newblock \doi{10.1016/j.physletb.2012.08.021}

\bibitem{Kajantie:1996mn}
K.~Kajantie, M.~Laine, K.~Rummukainen, M.E. Shaposhnikov, Phys. Rev. Lett.
  \textbf{77}, 2887 (1996).
\newblock \doi{10.1103/PhysRevLett.77.2887}

\bibitem{Kajantie:1996qd}
K.~Kajantie, M.~Laine, K.~Rummukainen, M.E. Shaposhnikov, Nucl. Phys. B
  \textbf{493}, 413 (1997).
\newblock \doi{10.1016/S0550-3213(97)00164-8}

\bibitem{Mazumdar:2018dfl}
A.~Mazumdar, G.~White, Rept. Prog. Phys. \textbf{82}(7), 076901 (2019).
\newblock \doi{10.1088/1361-6633/ab1f55}

\bibitem{Hashino:2018wee}
K.~Hashino, R.~Jinno, M.~Kakizaki, S.~Kanemura, T.~Takahashi, M.~Takimoto,
  Phys. Rev. D \textbf{99}(7), 075011 (2019).
\newblock \doi{10.1103/PhysRevD.99.075011}

\bibitem{Alves:2018jsw}
A.~Alves, T.~Ghosh, H.K. Guo, K.~Sinha, D.~Vagie, JHEP \textbf{04}, 052 (2019).
\newblock \doi{10.1007/JHEP04(2019)052}

\bibitem{Kurup:2017dzf}
G.~Kurup, M.~Perelstein, Phys. Rev. D \textbf{96}(1), 015036 (2017).
\newblock \doi{10.1103/PhysRevD.96.015036}

\bibitem{Hashino:2016xoj}
K.~Hashino, M.~Kakizaki, S.~Kanemura, P.~Ko, T.~Matsui, Phys. Lett. B
  \textbf{766}, 49 (2017).
\newblock \doi{10.1016/j.physletb.2016.12.052}

\bibitem{Kakizaki:2015wua}
M.~Kakizaki, S.~Kanemura, T.~Matsui, Phys. Rev. D \textbf{92}(11), 115007
  (2015).
\newblock \doi{10.1103/PhysRevD.92.115007}

\bibitem{Weinberg:1975gm}
S.~Weinberg, Phys. Rev. D \textbf{13}, 974 (1976).
\newblock \doi{10.1103/PhysRevD.19.1277}.
\newblock [Addendum: Phys.Rev.D 19, 1277--1280 (1979)]

\bibitem{Susskind:1978ms}
L.~Susskind, Phys. Rev. D \textbf{20}, 2619 (1979).
\newblock \doi{10.1103/PhysRevD.20.2619}

\bibitem{Peskin:1990zt}
M.E. Peskin, T.~Takeuchi, Phys. Rev. Lett. \textbf{65}, 964 (1990).
\newblock \doi{10.1103/PhysRevLett.65.964}

\bibitem{Peskin:1991sw}
M.E. Peskin, T.~Takeuchi, Phys. Rev. D \textbf{46}, 381 (1992).
\newblock \doi{10.1103/PhysRevD.46.381}

\bibitem{Appelquist:1986an}
T.W. Appelquist, D.~Karabali, L.C.R. Wijewardhana, Phys. Rev. Lett.
  \textbf{57}, 957 (1986).
\newblock \doi{10.1103/PhysRevLett.57.957}

\bibitem{Foadi:2007ue}
R.~Foadi, M.T. Frandsen, T.A. Ryttov, F.~Sannino, Phys. Rev. D \textbf{76},
  055005 (2007).
\newblock \doi{10.1103/PhysRevD.76.055005}

\bibitem{Simmons:1988fu}
E.H. Simmons, Nucl. Phys. B \textbf{312}, 253 (1989).
\newblock \doi{10.1016/0550-3213(89)90296-4}

\bibitem{Kagan:1991gh}
A.~Kagan, S.~Samuel, Phys. Lett. B \textbf{270}, 37 (1991).
\newblock \doi{10.1016/0370-2693(91)91535-4}

\bibitem{Jarvinen:2009mh}
M.~Jarvinen, C.~Kouvaris, F.~Sannino, Phys. Rev. D \textbf{81}, 064027 (2010).
\newblock \doi{10.1103/PhysRevD.81.064027}

\bibitem{Jarvinen:2010ms}
M.~Jarvinen, J. Phys. Conf. Ser. \textbf{259}, 012053 (2010).
\newblock \doi{10.1088/1742-6596/259/1/012053}

\bibitem{Miura:2018dsy}
K.~Miura, H.~Ohki, S.~Otani, K.~Yamawaki, JHEP \textbf{10}, 194 (2019).
\newblock \doi{10.1007/JHEP10(2019)194}

\bibitem{Azatov:2020nbe}
A.~Azatov, M.~Vanvlasselaer, JHEP \textbf{09}, 085 (2020).
\newblock \doi{10.1007/JHEP09(2020)085}

\bibitem{Pasechnik:2013bxa}
R.~Pasechnik, V.~Beylin, V.~Kuksa, G.~Vereshkov, Phys. Rev. D \textbf{88}(7),
  075009 (2013).
\newblock \doi{10.1103/PhysRevD.88.075009}

\bibitem{Lee:1967ug}
B.W. Lee, H.T. Nieh, Phys. Rev. \textbf{166}, 1507 (1968).
\newblock \doi{10.1103/PhysRev.166.1507}

\bibitem{Gasiorowicz:1969kn}
S.~Gasiorowicz, D.A. Geffen, Rev. Mod. Phys. \textbf{41}, 531 (1969).
\newblock \doi{10.1103/RevModPhys.41.531}

\bibitem{Ko:1994en}
P.~Ko, S.~Rudaz, Phys. Rev. D \textbf{50}, 6877 (1994).
\newblock \doi{10.1103/PhysRevD.50.6877}

\bibitem{Urban:2001ru}
M.~Urban, M.~Buballa, J.~Wambach, Nucl. Phys. A \textbf{697}, 338 (2002).
\newblock \doi{10.1016/S0375-9474(01)01248-9}

\bibitem{Pasechnik:2014ida}
R.~Pasechnik, V.~Beylin, V.~Kuksa, G.~Vereshkov, Int. J. Mod. Phys. A
  \textbf{31}(08), 1650036 (2016).
\newblock \doi{10.1142/S0217751X16500366}

\bibitem{Coleman:1973jx}
S.R. Coleman, E.J. Weinberg, Phys. Rev. D \textbf{7}, 1888 (1973).
\newblock \doi{10.1103/PhysRevD.7.1888}

\bibitem{Dolan:1973qd}
L.~Dolan, R.~Jackiw, Phys. Rev. D \textbf{9}, 3320 (1974).
\newblock \doi{10.1103/PhysRevD.9.3320}

\bibitem{Quiros:1999jp}
M.~Quiros, in \emph{{ICTP Summer School in High-Energy Physics and Cosmology}}
  (1999), pp. 187--259

\bibitem{Parwani:1991gq}
R.R. Parwani, Phys. Rev. D \textbf{45}, 4695 (1992).
\newblock \doi{10.1103/PhysRevD.45.4695}.
\newblock [Erratum: Phys.Rev.D 48, 5965 (1993)]

\bibitem{Arnold:1992rz}
P.B. Arnold, O.~Espinosa, Phys. Rev. D \textbf{47}, 3546 (1993).
\newblock \doi{10.1103/PhysRevD.47.3546}.
\newblock [Erratum: Phys.Rev.D 50, 6662 (1994)]

\bibitem{Gerber:1988tt}
P.~Gerber, H.~Leutwyler, Nucl. Phys. B \textbf{321}, 387 (1989).
\newblock \doi{10.1016/0550-3213(89)90349-0}

\bibitem{Ioffe:2001bn}
B.L. Ioffe, Phys. Usp. \textbf{44}, 1211 (2001).
\newblock \doi{10.1070/PU2001v044n12ABEH000972}

\bibitem{Hindmarsh:2015qta}
M.~Hindmarsh, S.J. Huber, K.~Rummukainen, D.J. Weir, Phys. Rev. D
  \textbf{92}(12), 123009 (2015).
\newblock \doi{10.1103/PhysRevD.92.123009}

\bibitem{Hindmarsh:2017gnf}
M.~Hindmarsh, S.J. Huber, K.~Rummukainen, D.J. Weir, Phys. Rev. D
  \textbf{96}(10), 103520 (2017).
\newblock \doi{10.1103/PhysRevD.96.103520}.
\newblock [Erratum: Phys.Rev.D 101, 089902 (2020)]

\bibitem{Grojean:2006bp}
C.~Grojean, G.~Servant, Phys. Rev. D \textbf{75}, 043507 (2007).
\newblock \doi{10.1103/PhysRevD.75.043507}

\bibitem{Leitao:2015fmj}
L.~Leitao, A.~Megevand, JCAP \textbf{05}, 037 (2016).
\newblock \doi{10.1088/1475-7516/2016/05/037}

\bibitem{Caprini:2015zlo}
C.~Caprini, et~al., JCAP \textbf{04}, 001 (2016).
\newblock \doi{10.1088/1475-7516/2016/04/001}

\bibitem{Caprini:2019egz}
C.~Caprini, et~al., JCAP \textbf{03}, 024 (2020).
\newblock \doi{10.1088/1475-7516/2020/03/024}

\bibitem{Wang:2020jrd}
X.~Wang, F.P. Huang, X.~Zhang, JCAP \textbf{05}, 045 (2020).
\newblock \doi{10.1088/1475-7516/2020/05/045}

\bibitem{Coleman:1977py}
S.R. Coleman, Phys. Rev. D \textbf{15}, 2929 (1977).
\newblock \doi{10.1103/PhysRevD.16.1248}.
\newblock [Erratum: Phys.Rev.D 16, 1248 (1977)]

\bibitem{Callan:1977pt}
C.G. Callan, Jr., S.R. Coleman, Phys. Rev. D \textbf{16}, 1762 (1977).
\newblock \doi{10.1103/PhysRevD.16.1762}

\bibitem{Linde:1980tt}
A.D. Linde, Phys. Lett. B \textbf{100}, 37 (1981).
\newblock \doi{10.1016/0370-2693(81)90281-1}

\bibitem{Espinosa:2010hh}
J.R. Espinosa, T.~Konstandin, J.M. No, G.~Servant, JCAP \textbf{06}, 028
  (2010).
\newblock \doi{10.1088/1475-7516/2010/06/028}

\bibitem{Ellis:2019oqb}
J.~Ellis, M.~Lewicki, J.M. No, V.~Vaskonen, JCAP \textbf{06}, 024 (2019).
\newblock \doi{10.1088/1475-7516/2019/06/024}

\bibitem{Wainwright:2011kj}
C.L. Wainwright, Comput. Phys. Commun. \textbf{183}, 2006 (2012).
\newblock \doi{10.1016/j.cpc.2012.04.004}

\bibitem{Freitas:2021yng}
F.F. Freitas, G.~Louren\c{c}o, A.P. Morais, A.~Nunes, J.a. Ol\'\i{}via,
  R.~Pasechnik, R.~Santos, J.a. Viana, JCAP \textbf{03}(03), 046 (2022).
\newblock \doi{10.1088/1475-7516/2022/03/046}

\bibitem{Morais:2018uou}
A.P. Morais, R.~Pasechnik, T.~Vieu, PoS \textbf{EPS-HEP2019}, 054 (2020).
\newblock \doi{10.22323/1.364.0054}

\bibitem{Morais:2019fnm}
A.P. Morais, R.~Pasechnik, JCAP \textbf{04}, 036 (2020).
\newblock \doi{10.1088/1475-7516/2020/04/036}

\bibitem{Greljo:2019xan}
A.~Greljo, T.~Opferkuch, B.A. Stefanek, Phys. Rev. Lett. \textbf{124}(17),
  171802 (2020).
\newblock \doi{10.1103/PhysRevLett.124.171802}

\bibitem{Aoki:2021oez}
M.~Aoki, T.~Komatsu, H.~Shibuya,   (2021)

\bibitem{Zhou:2020stj}
Z.~Zhou, J.~Yan, A.~Addazi, Y.F. Cai, A.~Marciano, R.~Pasechnik, Phys. Lett. B
  \textbf{812}, 136026 (2021).
\newblock \doi{10.1016/j.physletb.2020.136026}

\bibitem{Schmitz:2020syl}
K.~Schmitz, JHEP \textbf{01}, 097 (2021).
\newblock \doi{10.1007/JHEP01(2021)097}

\bibitem{LISA:2017pwj}
P.~Amaro-Seoane, et~al.,   (2017)

\bibitem{Corbin:2005ny}
V.~Corbin, N.J. Cornish, Class. Quant. Grav. \textbf{23}, 2435 (2006).
\newblock \doi{10.1088/0264-9381/23/7/014}

\bibitem{Kudoh:2005as}
H.~Kudoh, A.~Taruya, T.~Hiramatsu, Y.~Himemoto, Phys. Rev. D \textbf{73},
  064006 (2006).
\newblock \doi{10.1103/PhysRevD.73.064006}

\bibitem{Gowling:2021gcy}
C.~Gowling, M.~Hindmarsh, JCAP \textbf{10}, 039 (2021).
\newblock \doi{10.1088/1475-7516/2021/10/039}

\bibitem{Croon:2020cgk}
D.~Croon, O.~Gould, P.~Schicho, T.V.I. Tenkanen, G.~White, JHEP \textbf{04},
  055 (2021).
\newblock \doi{10.1007/JHEP04(2021)055}

\bibitem{Gould:2021oba}
O.~Gould, T.V.I. Tenkanen, JHEP \textbf{06}, 069 (2021).
\newblock \doi{10.1007/JHEP06(2021)069}

\bibitem{Camargo-Molina:2016moz}
J.E. Camargo-Molina, A.P. Morais, R.~Pasechnik, M.O.P. Sampaio, J.~Wess\'en,
  JHEP \textbf{08}, 073 (2016).
\newblock \doi{10.1007/JHEP08(2016)073}

\end{thebibliography}

\end{document}